\documentclass[reprint,
 amsmath,amssymb,
 aps,
]{revtex4-2}

\usepackage{graphicx}
\usepackage{dcolumn}
\usepackage{bm}
\usepackage{xcolor}

\begin{document}
\title{Temporal Talbot effect in free space}

\author{Layton A. Hall$^{1}$}
\author{Sergey A. Ponomarenko$^{2,3}$}
\author{Ayman F. Abouraddy$^{1}$}
\affiliation{$^{1}$CREOL, The College of Optics \& Photonics, University of Central~Florida, Orlando, FL 32816, USA}
\affiliation{$^{2}$Department of Electrical and Computer Engineering, Dalhousie University, Halifax, Nova Scotia B3J 2X4, Canada}%
\affiliation{$^{3}$Department of Physics and Atmospheric Science, Dalhousie University, Halifax, Nova Scotia B3H 4R2, Canada}%

\begin{abstract}
The temporal Talbot effect refers to the periodic revivals of a pulse train propagating in a dispersive medium, and is a temporal analog of the spatial Talbot effect with group-velocity dispersion in time replacing diffraction in space. Because of typically large temporal Talbot lengths, this effect has been observed to date in only single-mode fibers, rather than with freely propagating fields in bulk dispersive media. Here we demonstrate for the first time the temporal Talbot effect \textit{in free space} by employing dispersive space-time wave packets, whose spatio-temporal structure induces group-velocity dispersion of controllable magnitude and sign in free space. 
\end{abstract}

\maketitle

The Talbot effect, reported for the first time in 1836 \cite{Talbot36PM}, refers to the axial revivals of an initially periodic transverse spatial field structure \cite{Wen13AOP}. This fascinating phenomenon has found a broad range of applications, spanning structured illumination in fluorescence microscopy \cite{Han13AC,Sun16JBO,Chowdhury18arxiv} to prime-number decomposition \cite{Pelka18OE}, and phase-locking of laser
arrays \cite{Tradonsky17AO}. In an analogous \textit{temporal} Talbot effect, whereupon group velocity dispersion (GVD) in time replaces diffraction in space \cite{Kolner94IEEEJQE,Mitschke98OPN}, a periodic pulse train of period $T$ traveling in a dispersive medium undergoes periodic revivals at multiples of the temporal Talbot distance $z_{\mathrm{T}}\!=\!\tfrac{T^{2}}{\pi|k_{2}|}$, where $k_{2}$ is the GVD parameter \cite{Mitschke98OPN}. This effect was proposed in \cite{Jannson81JOSA}, demonstrated experimentally in \cite{Andrekson93OL} (and subsequently in \cite{Arahira98JLT,Shake98EL}), and has been exploited in removing pulse distortion \cite{Azana99AO} and in pulse-rate multiplication \cite{Arahira98JLT,Atkins03IEEE}.

\begin{figure}[t!]
\centering
\includegraphics[width=8.6cm]{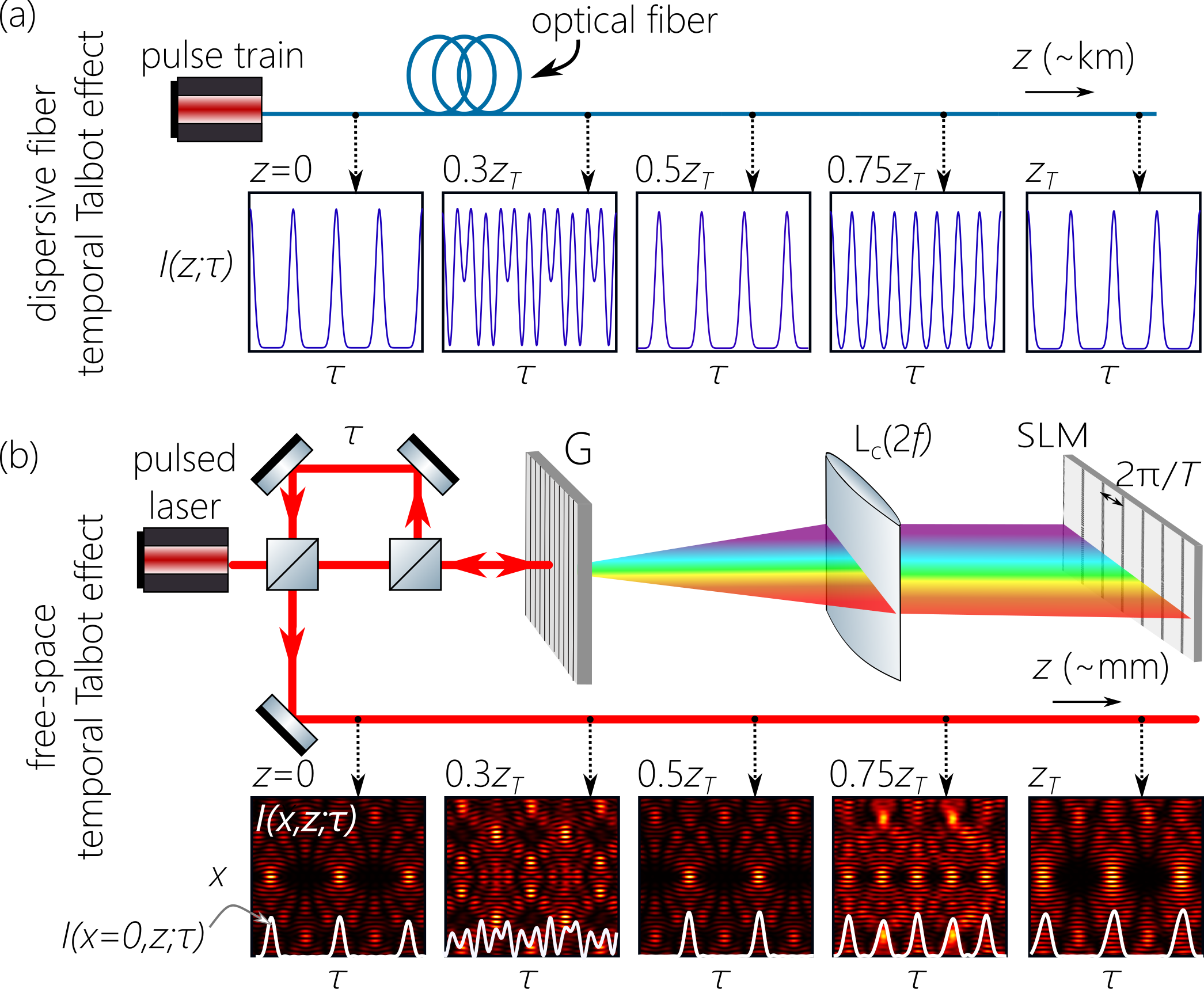}
\caption{(a) The temporal Talbot effect for a periodic pulse train manifested in axial revivals along a dispersive optical fiber. The plots are the intensity $I(z;\tau)$ along $z$; $z_{\mathrm{T}}$ is the temporal Talbot length. (b) The temporal Talbot effect realized in free space via dispersive ST wave packets. Schematic of the setup; G: diffraction grating, L$_{\mathrm{c}}$: cylindrical lens, SLM: spatial light modulator. The panels display the spatio-temporal intensity $I(x,z;\tau)$ at different $z$, and the white curves are the on-axis profiles $I(0,z;\tau)$, which are identical to $I(z;\tau)$ in (a).}
\label{Fig:Concept}
\end{figure}

The temporal Talbot effect has yet to be observed in a freely propagating optical field. Because dispersion lengths for typical pulse trains is usually very large, the temporal Talbot effect has been instead realized only in single-mode fibers ($z_{\mathrm{T}}$ on the order of kilometers, with $k_{2}\!\approx\!-26$~fs$^2$/mm at 1500~nm) \cite{Andrekson93OL}, or in fiber Bragg gratings with higher GVD \cite{Azana99AO} ($z_{\mathrm{T}}$ on the order of tens of centimeters with $k_{2}\!\approx\!-10^{5}$~fs$^2$/mm) [Fig.~\ref{Fig:Concept}(a)], but \textit{not} in dispersive bulk media where diffraction that unavoidably accompanies propagation hampers its observation.

\begin{figure*}[t!]
\centering
\includegraphics[width=17.6cm]{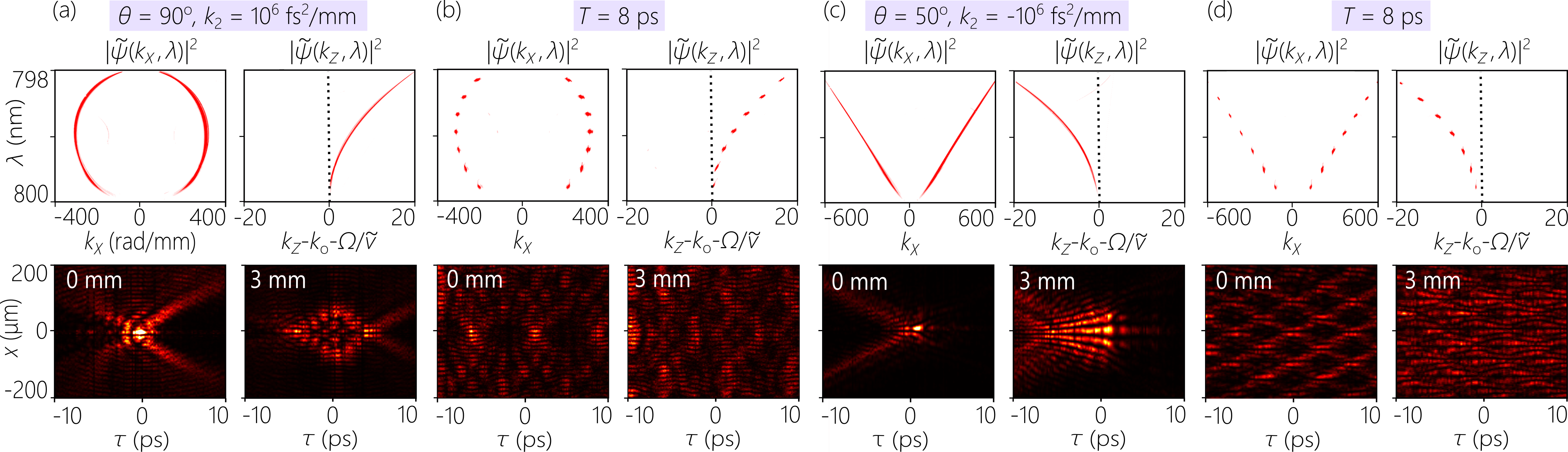}
\caption{First row shows continuous and distcretized spectral projections onto the $(k_{x},\lambda)$ and $(k_{z},\lambda)$ planes, $|\widetilde{\psi}(k_{x},\lambda)|^{2}$ and $|\widetilde{\psi}(k_{z},\lambda)|^{2}$, respectively, for dispersive ST wave packets. The second row shows the spatio-temporal intensity profiles at $z\!=\!0$ and $z\!=\!3$~mm for each wave packet. The dotted vertical line in the spectral projection onto the $(k_{z},\lambda)$-plane corresponds to a GVD-free ST wave packet. (a) Dispersive ST wave packet having $\theta\!=\!50^{\circ}$ and normal $k_{2}\!=\!1000k_{2}^{\mathrm{ZnSe}}$. (b) Same as (a) after discretizing the spectrum to produce a pulse train of period $T\!=\!8$~ps. (c,d) Same as (a,b), except for $\theta\!=\!90^{\circ}$ and anomalous GVD $k_{2}\!=\!-1000k_{2}^{\mathrm{ZnSe}}$.}
\label{Fig:Spectra}
\end{figure*}

Here we demonstrate -- for the first time to the best of our knowledge -- the temporal Talbot effect in a freely propagating field over short distances (a few centimeters) \textit{in free space}, without resort to any dispersive medium [Fig.~\ref{Fig:Concept}(b)]. This surprising effect is made possible by exploiting dispersive `space-time' (ST) wave packets \cite{Yessenov21arxiv}. In general, ST wave packets \cite{Kondakci16OE,Parker16OE,Yessenov19OPN} are pulsed beams endowed with a precise spatio-temporal structure \cite{Turunen10PO,Reivelt03arxiv,FigueroaBook14} inculcating angular dispersion \cite{Torres10AOP,Fulop10Review}, by virtue of which they display a variety of unique behaviors, including propagation invariance \cite{Kondakci17NP,Kondakci18PRL,Bhaduri19OL,Schepler20ACSP,Shiri20NC,Shiri20OL}; tunable group velocities in absence of dispersion \cite{Kondakci19NC,Bhaduri19Optica,Bhaduri20NP}; self-healing \cite{Kondakci18OL}; free-space acceleration/deceleration \cite{Clerici08OE,Valtna09OE,Yessenov20PRL2,Li20CP}; among many other possibilities \cite{Wong17ACSP2,Shaltout19Science,Chong20NP,Wong20AS}. Rather than propagation-invariant ST wave packets, observing the temporal Talbot effect requires utilizing their counterparts exhibiting GVD in free space \cite{Yessenov21arxiv}. Because the angular dispersion underpinning ST wave packets is non-differentiable \cite{Hall21arxiv}, unlike conventional angular dispersion associated with tilted pulse fronts (TPFs) that is differentiable \cite{Torres10AOP,Fulop10Review}, ST wave packets can experience arbitrary GVD in free space \cite{Yessenov21arxiv}, whereas TPFs can experience only anomalous GVD \cite{Martinez84JOSAA,Torres10AOP,Fulop10Review}. After introducing normal or anomalous GVD of large magnitude, periodically sampling the temporal spectrum of the ST wave packet produces the temporal Talbot effect with $z_{\mathrm{T}}$ on the order of a few centimeters ($\approx\!2$~cm here). Crucially, because the spatial and temporal degrees of freedom are coupled, the initial (non-periodic) spatial profile is repeated at the temporal Talbot planes, thereby facilitating unambiguous observation of on-axis temporal revivals in free space for the first time.

We start by describing propagation-invariant ST wave packets in which each spatial frequency $k_{x}$ is associated with a single temporal frequency $\omega$ to ensure that the axial wave number $k_{z}$ is related linearly to $\omega$, $\Omega\!=\!(k_{z}-k_{\mathrm{o}})c\tan{\theta}$; here $\Omega\!=\!\omega-\omega_{\mathrm{o}}$ is the temporal frequency relative to a fixed frequency $\omega_{\mathrm{o}}$, $k_{\mathrm{o}}\!=\!\tfrac{\omega_{\mathrm{o}}}{c}$ is the corresponding wave number, $c$ is the speed of light in vacuum, $x$ and $z$ are the transverse and longitudinal coordinates, respectively, the field is held uniform along $y$ for simplicity, and we refer to $\theta$ as the spectral tilt angle. Geometrically, this construction is equivalent to restricting the spatio-temporal spectrum on the surface of the light-cone $k_{x}^{2}+k_{z}^{2}\!=\!(\tfrac{\omega}{c})^{2}$ to its intersection with a plane that is parallel to the $k_{x}$-axis and is tilted by an angle $\theta$ with respect to the $k_{z}$-axis, such that its projection onto the $(k_{z},\tfrac{\omega}{c})$-plane is the straight line $k_{z}\!=\!k_{\mathrm{o}}+\tfrac{\Omega}{c}\cot{\theta}$. Such a ST wave packet is propagation-invariant $\psi(x,z;t)\!=\!\psi(x,0;t-z/\widetilde{v})$, where $\psi(x,z;t)$ is the spatio-temporal envelope of the field $E(x,z;t)\!=\!e^{i(k_{\mathrm{o}}z-\omega_{\mathrm{o}}t)}\psi(x,z;t)$, and $\widetilde{v}\!=\!c\tan{\theta}$ is the group velocity \cite{Kondakci17NP}. By replacing the plane with a planar curved surface that is also parallel to the $k_{x}$-axis but whose projection onto the $(k_{z},\tfrac{\omega}{c})$-plane is the curve $k_{z}\!=\!k_{\mathrm{o}}+\Omega/\widetilde{v}+k_{2}\Omega^{2}/2$, then the envelope takes the form:
\begin{equation}\label{Eq:EnvelopeWithDiscretizedSpectrum}
\psi(x,z;t)=\int\!d\Omega\,\widetilde{\psi}(\Omega)e^{ik_{x}(\Omega)x}e^{-i\Omega(t-z/\widetilde{v})}e^{ik_{2}\Omega^{2}z/2}.
\end{equation}
The on-axis envelope $\psi(0,z;t)$ takes the form of a plane-wave pulse undergoing GVD (with GVD parameter $k_{2}$) along $z$, albeit in absence of a dispersive medium.

We introduce a periodic pulse train structure into the field by discretizing the temporal spectrum along $\omega$ at multiples of $\tfrac{2\pi}{T}$, $\Omega\!\rightarrow\!\Omega_{m}\!=\!m\tfrac{2\pi}{T}$ for integer $m$, so that the on-axis envelope is:
\begin{equation}
\psi(0,z;t)=\sum_{m}\widetilde{\psi}_{m}e^{-i2\pi m(t-z/\widetilde{v})/T}e^{i2\pi\,\mathrm{sgn}(k_{2})\,m^{2}z/z_{\mathrm{T}}},
\end{equation}
where $\mathrm{sgn}(k_{2})\!=\!\pm1$ is the sign of $k_{2}$, $\widetilde{\psi}_{m}\!=\!\widetilde{\psi}(\Omega_{m})$ and $z_{\mathrm{T}}\!=\! T^{2}/\pi|k_{2}|$. The tight association between temporal and spatial frequencies entails simultaneously discretizing the spatial spectrum along $k_{x}$. However, because $\omega$ and $k_{x}$ are \textit{not} linearly related, $k_{x}$ is therefore \textit{not} sampled periodically, and the transverse spatial profile at $z\!=\!0$ is thus not periodic. It is clear that the initial envelope is periodic in time $\psi(0,0;t+\ell T)\!=\!\psi(0,0;t)$, and undergoes axial revivals at the Talbot planes $\psi(0,\ell z_{\mathrm{T}};t)\!=\!\psi(0,0;t-\ell z/\widetilde{v})$ in a time frame traveling at $\widetilde{v}$.

We prepare the ST field using the 2D pulse synthesizer developed in Refs.~\cite{Kondakci17NP,Kondakci19NC,Bhaduri19Optica,Bhaduri20NP} and shown schematically in Fig.~\ref{Fig:Concept}(b). This arrangement implements a two-step spatio-temporal spectral synthesis strategy capable of producing arbitrary, non-differentiable angular dispersion \cite{Yessenov19OE,Kondakci19ACSP,Yessenov21arxiv}. Plane-wave pulses (pulsewidth $\sim\!100$~fs at a central wavelength $\sim\!800$~nm) from a mode-locked Ti:sapphire laser (Tsunami; Spectra Physics) are directed to a diffraction grating that spreads the pulse spectrum in space, whereupon the first diffraction order is collimated with a cylindrical lens before impinging on a reflective, phase-only spatial light modulator (SLM). The SLM imparts a 2D phase distribution to the spectrally resolved wave front that assigns to each wavelength $\lambda$ a spatial frequency $k_{x}(\lambda)$ to guarantee that $k_{z}(\Omega,k_{x})\!=\!k_{\mathrm{o}}+\Omega/\widetilde{v}+k_{2}\Omega^{2}/2$, for given $\widetilde{v}$ and $k_{2}$. The retro-reflected field returns to the grating whereupon the ST wave packets are reconstituted with an on-axis pulsewidth of $\approx\!2$~ps. We measure the spatio-temporal spectrum via a combination of a grating and a lens to carry out temporal and spatial Fourier transforms, and we obtian the spatio-temporal intensity profile by interfering the ST field with a short plane-wave reference pulse obtained directly from the Ti:sapphire laser \cite{Kondakci19NC,Bhaduri20NP}.

\begin{figure}[t!]
\centering
\includegraphics[width=8.6cm]{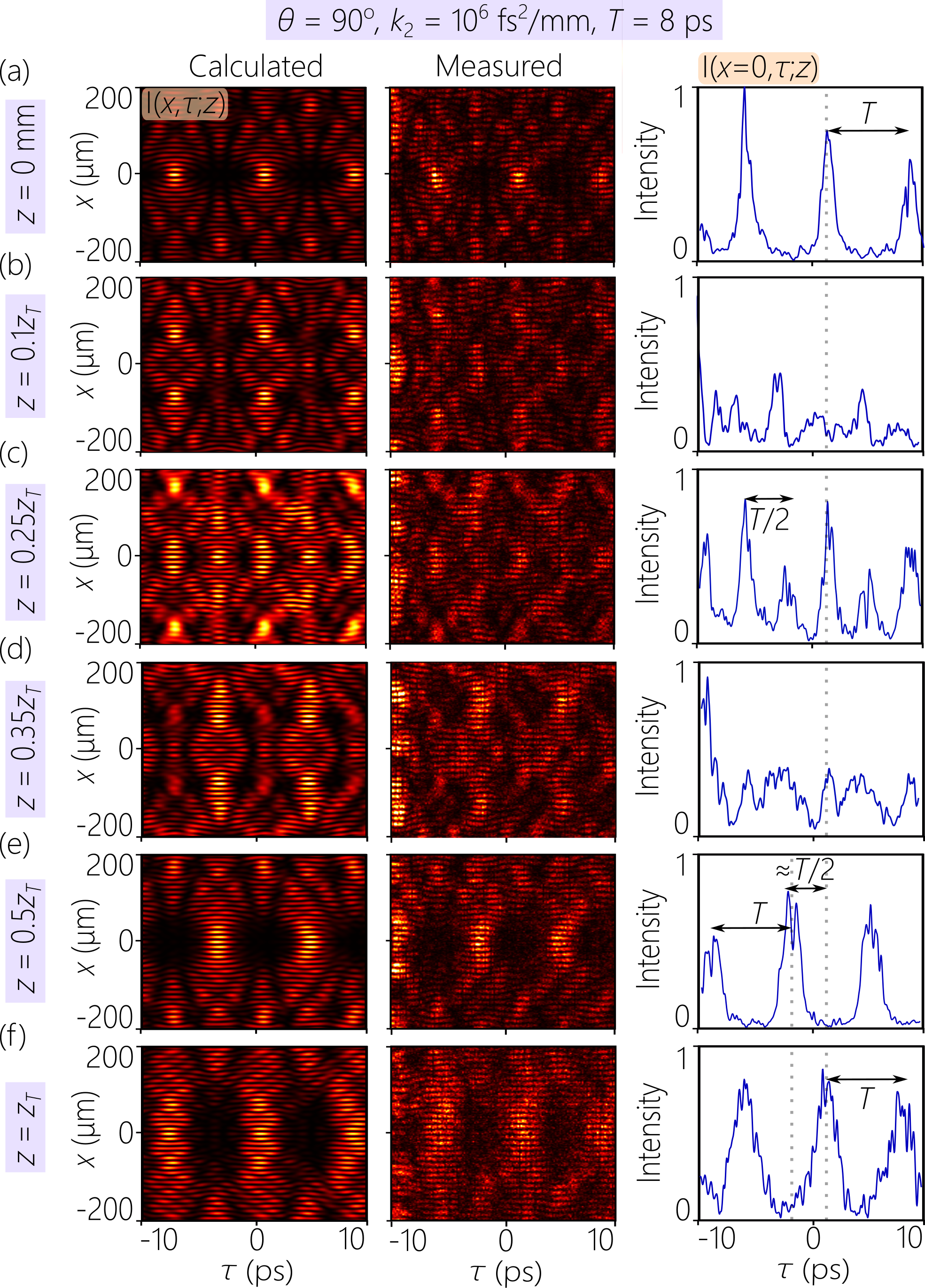}
\caption{Demonstration of the temporal Talbot effect in free space employing the dispersive ST wave packet experiencing normal dispersion in free space from Fig.~\ref{Fig:Spectra}(b).}
\label{Fig:Measurements}
\end{figure}

\begin{figure}[t!]
\centering
\includegraphics[width=8.6cm]{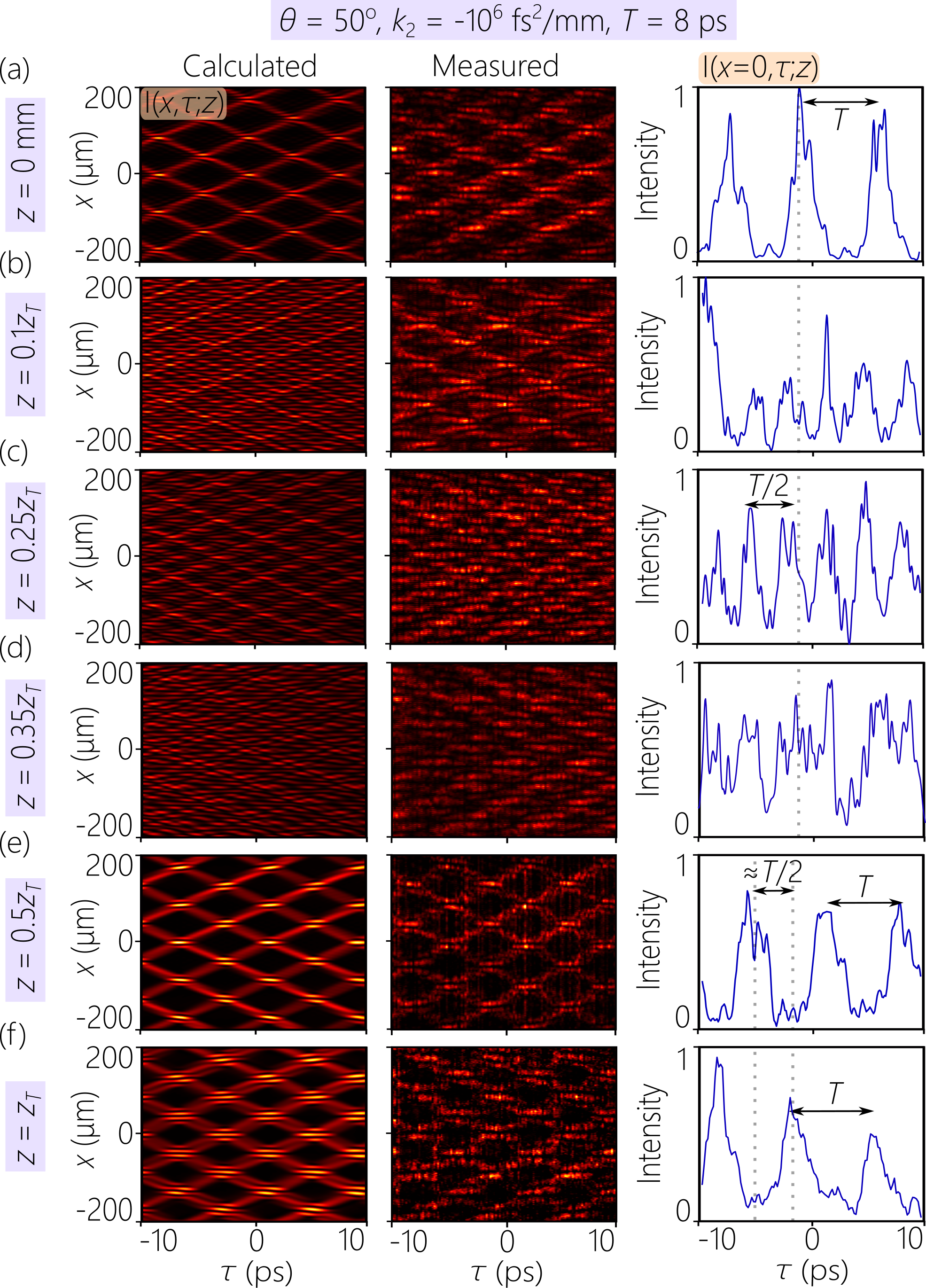}
\caption{Demonstration of the temporal Talbot effect in free space employing the dispersive ST wave packet experiencing anomalous dispersion in free space from Fig.~\ref{Fig:Spectra}(d).}
\label{Fig:Measurements2}
\end{figure}

We plot in Fig.~\ref{Fig:Spectra} the measured spatio-temporal spectra for dispersive ST wave packets. In Fig.~\ref{Fig:Spectra}(a) we plot the spectral projections onto the $(k_{x},\lambda)$ and $(k_{z},\lambda)$ planes for normal GVD whereupon the spectral projection onto the $(k_{x},\lambda)$-plane is O-shaped. The GVD parameter here is $k_{2}\!=\!10^6$~fs$^2$/mm, which is significantly larger than ZnSe $k_{2}\!\approx\!10^{3}$~fs$^2$/mm (at $\lambda\!=\!800$~nm) and Bragg gratings ($k_{2}\!\approx\!10^{5}$~fs$^2$/mm). Once the temporal spectrum is discretized [Fig.~\ref{Fig:Spectra}(b)] to produce a period $T\!=\!8$~ps, an on-axis ($x\!=\!0$) periodic pulse-train structure emerges with a predicted temporal Talbot length of $z_{\mathrm{T}}\approx\!20$~mm because of the rapidly dispersing wave packet. We plot in Fig.~\ref{Fig:Spectra}(c) the measured spatio-temporal spectral projections onto the $(k_{x},\lambda)$ and $(k_{z},\lambda)$ planes after introducing anomalous GVD equal in magnitude but opposite in sign to that in Fig.~\ref{Fig:Spectra}(a). In absence of GVD, the projection onto the $(k_{x},\lambda)$ is a parabola, but becomes V-shaped in presence of large anomalous GVD. We then plot in Fig.~\ref{Fig:Spectra}(d) the spectral projections and intensity profiles after spectral discretization with $T\!=\!8$~ps. 

\begin{figure}[t!]
\centering
\includegraphics[width=8.6cm]{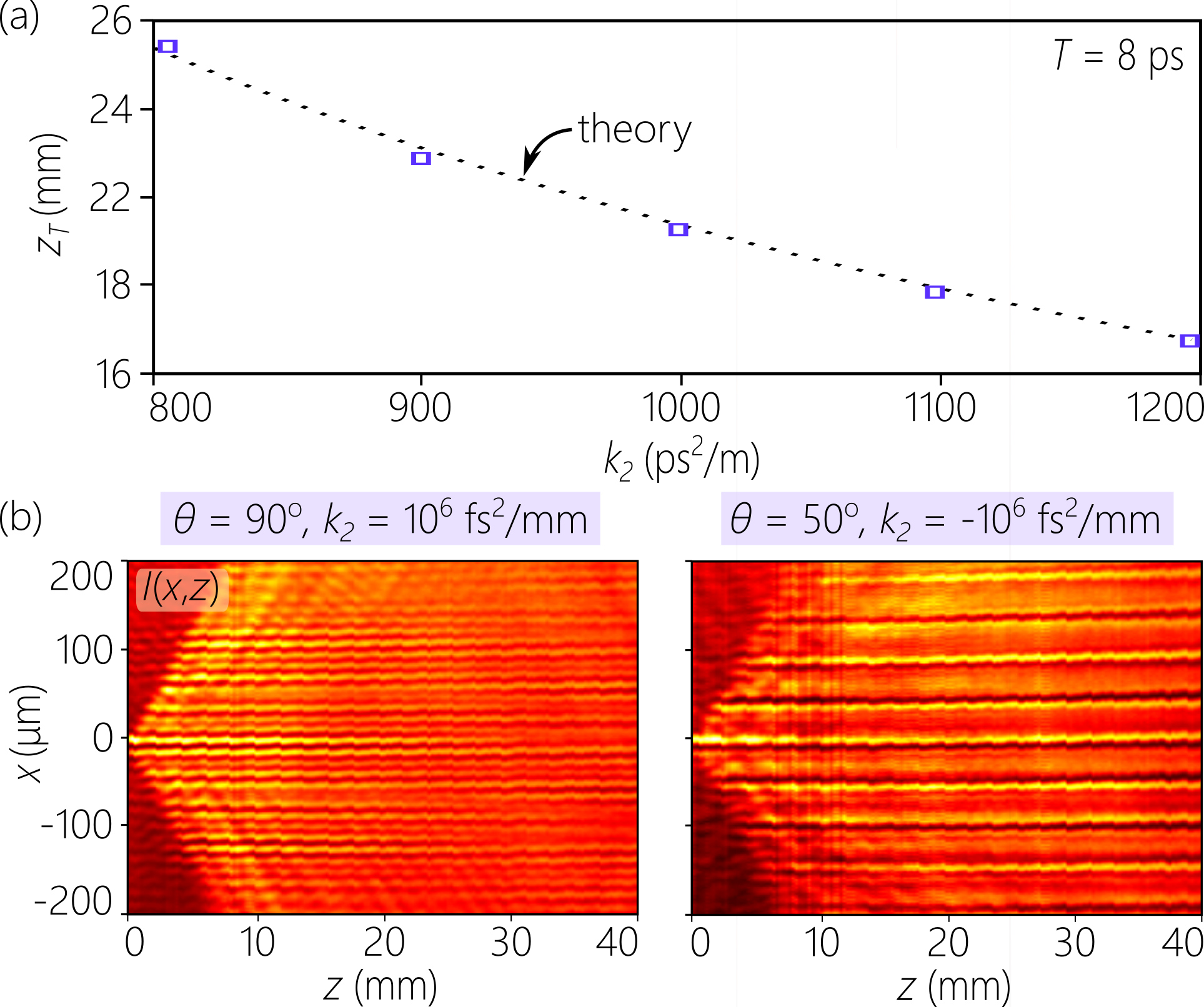}
\caption{(a) Measured temporal Talbot length $z_{\mathrm{T}}$ while varying the GVD parameter in the normal regime. The dotted curve is $z_{\mathrm{T}}\!=\!T^{2}/(\pi|k_{2}|)$. (b) The time-averaged intensity $I(x,z)$ for both normal and anomalous GVD, showing axial invariance despite the underlying temporal evolution (Fig.~\ref{Fig:Measurements} and Fig.~\ref{Fig:Measurements2}).}
\label{Fig:DependenceOnGVD}
\end{figure}

Despite the clear distinction between the profiles for normally dispersive ST fields with continuous and discretized spectra [Fig.~\ref{Fig:Spectra}(a,b)] and their anomalously dispersive counterparts [Fig.~\ref{Fig:Spectra}(c,d)], the on-axis intensity in both are nevertheless similar (Eq.~\ref{Eq:EnvelopeWithDiscretizedSpectrum}) with both exhibiting axial revivals of the initial periodic temporal profile. The measurement results for axial propagation of the dispersive ST wave packets alongside theoretical predictions are presented in Fig.~\ref{Fig:Measurements} for normal GVD corresponding to Fig.~\ref{Fig:Spectra}(b), and in Fig.~\ref{Fig:Measurements2} for anomalous GVD corresponding to Fig.~\ref{Fig:Spectra}(d), both with $T\!=\!8$~ps. We measure the temporally resolved intensity at the axial planes $z\!=\!0.5z_{\mathrm{T}}$, $0.6z_{\mathrm{T}}$, $0.75z_{\mathrm{T}}$, $0.85z_{\mathrm{T}}$, $z_{\mathrm{T}}$, and $1.5z_{\mathrm{T}}$. There is excellent agreement between the calculated (first column) and measured (second column) intensity profiles. The on-axis temporal profiles (third column)  reveal several critical features. First, the initial period profile [Fig.~\ref{Fig:Measurements}(a) and Fig.~\ref{Fig:Measurements2}(a)] is retrieved at the Talbot planes $z\!=\!mz_{\mathrm{T}}$ [Fig.~\ref{Fig:Measurements}(f) and Fig.~\ref{Fig:Measurements2}(f)].
Second, the periodic profile is reconstructed at the Talbot half-planes $z\!=\!(m+\frac{1}{2})z_{\mathrm{T}}$ but with a temporal displacement by $T/2$ with respect to $z\!=\!mz_{\mathrm{T}}$ [Fig.~\ref{Fig:Measurements}(e) and Fig.~\ref{Fig:Measurements2}(e)] . Third, at $z\!=\!0.25z_{\mathrm{T}}$ a rate doubling is observed; i.e., a periodic profile is observed but with period $T/2$ rather than $T$ [Fig.~\ref{Fig:Measurements}(c) and Fig.~\ref{Fig:Measurements2}(c)] . We repeat the measurements for different values of the GVD parameter $k_{2}$ and obtain the temporal Talbot length $z_{\mathrm{T}}$. The data plotted in Fig.~\ref{Fig:DependenceOnGVD}(a) shows excellent agreement with the theoretical expectation of $z_{\mathrm{T}}\!=\!\tfrac{T^{2}}{\pi|k_{2}|}$ with $T\!=\!8$~ps.

We recently reported a phenomenon we denoted the `veiled' Talbot effect resulting from periodically sampling the spatial spectrum along $k_{x}$ for a \textit{propagation-invariant} ST wave packet \cite{Yessenov20PRL1}. The conventional spatial Talbot effect was observed in time-resolved measurements as a consequence of time diffraction \cite{Moshinsky52PR,Longhi04OE,Porras17OL,Kondakci18PRL}, but no temporal dynamics are observed in absence of GVD. The time-averaged intensity (or energy) is diffraction-free along $z$ with a period $L/2$ rather than $L$. In the work reported here, the transverse profile is \textit{not} periodic, and yet the time-averaged intensity remains diffraction-free, as shown in the measurements plotted in Fig.~\ref{Fig:DependenceOnGVD}(b) for normal and anomalous GVD, despite the underlying axial dynamics (Fig.~\ref{Fig:Measurements} and Fig.~\ref{Fig:Measurements2}).

We have also reported on a ST Talbot effect based on the unique dispersive ST wave packet  denoted a `V-wave' whose diffraction and dispersion lengths are intrinsically equal \cite{Hall21arxivTalbot}. Because $k_{x}$ and $\omega$ are linearly related in a V-wave, $k_{x}$ and $\omega$ can be simultaneously sampled periodically to guarantee equal spatial and temporal Talbot lengths. However, this is a restrictive condition that admits of this unique solution. The Talbot effect we present here is purely temporal, and a broad range of dispersive ST wave packets can be utilized in realizing it. Crucially, V-waves are endowed with differentiable angular dispersion, so they are amenable to the conventional perturbative theory \cite{Martinez84JOSAA} and they therefore can inculcate only anomalous GVD. In contrast, non-differentiable angular dispersion is introduced in the the dispersive ST fields examined here \cite{Hall21arxiv} and they can undergo either normal or anomalous GVD \cite{Yessenov21arxiv}.

In conclusion, we have observed for the first time the temporal Talbot effect with a freely propagating field rather than a confined mode in a single-mode fiber. This demonstration made use of dispersive ST wave packets undergirded by non-differentiable angular dispersion that allows us to induce in free space normal or anomalous GVD of extremely large magnitudes, thus reducing the temporal Talbot length to $z_{\mathrm{T}}\!\sim\!20$~mm. Moreover, the initial non-periodic \textit{spatial} profile is simultaneously revived at the \textit{temporal} Talbot planes, thereby allowing for the unambiguous observation of this effect.

\section*{Funding}
U.S. Office of Naval Research (ONR) contract N00014-17-1-2458 and ONR MURI contract N00014-20-1-2789.

\section*{Disclosures}
The authors declare no conflicts of interest.

\bibliography{diffraction}

\end{document}